\documentclass[5p, times]{elsarticle}

\makeatletter
\def\ps@pprintTitle{%
 \let\@oddhead\@empty
 \let\@evenhead\@empty
 \def\@oddfoot{}%
 \let\@evenfoot\@oddfoot}
\makeatother

\usepackage{graphicx}
\usepackage{amssymb}
\usepackage{amsthm}
\usepackage{amsmath}
\usepackage{float}
\usepackage{subcaption}

\usepackage{verbatim}

\usepackage{xcolor}

\usepackage{lineno}

\biboptions{comma,round,authoryear}

\begin{document}

\begin{frontmatter}


\title{Variety, Complexity and Economic Development}


\author{Alje van Dam \fnref{label1,label2}}
\ead{A.vanDam@uu.nl}

\author{Koen Frenken \fnref{label1}\corref{cor1}}
\ead{K.Frenken@uu.nl}


\fntext[label1]{Copernicus Institute of Sustainable Development, Utrecht University}

\fntext[label2]{Center for Complex Systems Studies, Utrecht University}

\cortext[cor1]{Corresponding author}


%%\address{California, United States}

\begin{abstract}
We propose a combinatorial model of economic development. An economy develops by acquiring new capabilities allowing for the production of an ever greater variety of products of increasingly complex products. Taking into account that economies abandon the least complex products as they develop over time, we show that variety first increases and then decreases in the course of economic development. This is consistent with the empirical pattern known as 'the hump'. Our results question the common association of variety with complexity. We further discuss the implications of our model for future research.
\end{abstract}


\end{frontmatter}

%\linenumbers
\newpage

\section{Introduction}
\label{S:1} 
Our understanding of economic growth has long been guided by the notion of a production function that specifies how inputs such as capital and labour translate into the total output of an economy. Theoretical models of economic growth typically abstracted from the exact products that an economy produces, describing economic growth instead as an increase in aggregate output.

%complementarity 
Recently, more attention has been given to the specific products an economy produces \citep{Hausmann2007}. At the level of products, inputs can be considered to be strictly complementary \citep{Kremer1993, Hausmann2011, Brummitt2017}. This assumption is based on the idea that the production of any product or service requires a particular combination of complementary inputs. Missing one of those inputs renders the others useless in the production process.

%capabilities 
Inputs required to produce a product can be many, and include physical resources and assets as well as knowledge, skills, and even regulations. All these inputs are often referred to in an abstract and generic sense as 'capabilities' \citep{Hidalgo2009, Hausmann2011}. Products can then be represented as strings of capabilities. The ability of an economy to produce products (including services) thus depends on the number of capabilities present in a country, as well as the ways in which capabilities complement each other.

%diversification
Developing new products consists of recombining old and new inputs into configurations that have economic value \citep{Inoua2016}. Since these new recombinations will consist largely of capabilities that were already present, new products will be similar, or 'related', to existing ones. The process of development can thus be described as one that is exploring new products in the 'adjacent possible' of the current set of capabilities \citep{Kauffman1993}. This implies that economic development is a highly path-dependent process \citep{Lall2000} characterized by a logic of related diversification \citep{Hidalgo2007}.

%sophistication 
The acquisition of new capabilities does not only allow an economy to increase its variety of products, but also more complex products in terms of the number of capabilities used in products. Combining more capabilities implies a more intricate production process leading to products that are arguably more sophisticated compared to combining only few capabilities. This line of thinking is consistent with the notion of product sophistication, and the idea that sophisticated products can only be produced in well-developed economics with many capabilities \citep{Lall2006, Hausmann2007, Sutton2016}.

%Two strands of research have emerged from this combinatorial framework. 
Two streams of research have followed from this combinatorial framework. First, empirical studies have investigated the role of relatedness in economic development. New products will be related to existing products in that new products are produced using both existing and newly acquired capabilities. Following this reasoning, studies have analysed the extent to which national and regional economies diversify over time from existing products into related products \citep{Hidalgo2007, Neffke2011}.

Second, there have been several attempts to measure the average complexity of products produced by a country. The proposed measures build on methods that infer the complexity of economies by iteratively weighing the variety of products produced in a country and the ubiquity of these products in other countries. Such indirect measures of complexity have been used to explain income differences across countries and their growth rates over time \citep{Hidalgo2009, Tacchella2012, Cristelli2015}.

%Gap 
Notwithstanding the explanatory power of aforementioned studies, the economic complexity framework so far neglects a salient and fundamental feature of economic development. While new products enter a country's portfolio as it develops, already existing products may also exit \citep{Cadot2011}. One reason that countries lose products from their portfolio holds that wages, over time, become so high that a country cannot remain competitive in certain products \citep{Sutton2016}. Products exiting the portfolio may thus be products that can be imported at lower prices from low-wage countries. In addition, some products may become obsolete once their functionality is substituted by new products. Either way, understanding economic development will logically have to take into account both products entering and products exiting at any moment in time.

%Hump
Empirically, it has been shown that the variety of products that an economy produces, is positively related to the income per capita of its workers \citep{Hesse2008, Herzer2006, Al-Marhubi2000}. This relationship, however, only holds up to a certain level of income per capita, as countries with the highest income per capita display lower variety. This inverted-U pattern between income per capita and variety is known as 'the hump' \citep{Imbs2003, Cadot2011}. In a dynamic sense, then, the hump suggests that in the course of development, countries first diversify and then specialize again. This empirical pattern is inconsistent with the basic model of economic complexity, which would predict an ever-increasing variety as more capabilities are acquired over time.

%contribution
We argue that products exiting a country's portfolio are likely to be the least complex ones. Such simple products can be imported at lower prices from low-income countries or substituted by new products entering a country's portfolio. Below, we extend the elementary combinatorial model underlying the framework of economic complexity by imposing a constraint on the range of the complexity of activities that an economy can engage in. As a result, at a certain stage of development, countries will start losing their least complex products. The introduction of this constraint results in a theoretical model that i. is consistent with the principle of related diversification, ii. recovers the stylized fact of 'the hump', and iii. predicts that the growth in economic complexity of an economy accelerates as a function of newly acquired capabilities. From the model, we further derive a number of research questions, in particular, regarding the nature of products exiting countries' portfolios and the variations across countries in terms of the timing of the hump. Finally, we will argue that, as our model suggests that complexity continues to increase while variety starts decreasing, empirical measures of complexity that rely on the measurement of variety are theoretically unsupported.

\section{A basic combinatorial model}
\label{S:2}
Following \cite{Inoua2016}, we start with a simple model in which every product is represented as a string of capabilities. The \textit{product length} is given by the number of capabilities required to produce it and indicates a product's sophistication or complexity.

The capabilities present in an economy determine the set of products that an economy can produce. For simplicity, we will assume that every possible combination of capabilities leads to a viable product (this assumption will be relaxed below where we introduce a 'recipe book'). A country that has $n$ capabilities can make ${n \choose s}$ different combinations of lengths $s$. The longest product it can produce is the one product that recombines all $n$ capabilities. The total number of products that a country can make is given by the total number of strings one can make out of $n$ capabilities
\begin{align*} 
d(n) = \sum_{s=0}^n {n \choose s} = 2^n. 
\end{align*} 
The average product length of those products is given by the total length of all products divided by the total number of products 
\begin{align*} 
\bar{s}(n) = \frac{\sum_{s=0}^n s {n \choose s}}{2^n} = \frac{n}{2}.
\end{align*} 
This leads to the following basic properties of the model:
\begin{enumerate}
    \item The product variety is given by $d(n) = 2^n$, so that $\log(d(n)) \propto n$. 
    \item The average product length in a country is given by $\bar{s}(n) = \frac{n}{2}$, so that $\bar{s}(n) \propto n$.
\end{enumerate}
Combining both properties, it follows that the logarithm of product variety is linearly proportional to the average product length: 
\begin{align*} 
\log(d) \propto \bar{s}(n).
\end{align*}
Further note that in this basic model both the logarithm of the product variety and the average product length in an economy could provide a measure of economic complexity, as they are both proportional to the number of capabilities present \citep{Inoua2016}. Furthermore, the exponential relation between product variety and the number of capabilities reflects that a country with many capabilities can increase its variety more by acquiring a new capability compared to a country with only few capabilities, since the former has more capabilities with which the new capability can be recombined than the latter. This aspect of the model has been associated with a 'poverty trap', as it predicts a divergence in the absolute levels of product variety for countries at different stages of development \citep{Hausmann2011}.

\section{Product exit}
\label{S:3}
We now extend the model to incorporate the possibility of an economy losing products. We pose that as the average complexity of products in a country keeps on rising as part of its economic development - and wages rise accordingly - it cannot remain competitive in the simplest products. As a result, a country will see its simplest products exit from its portfolio. This is modeled by imposing a product range $r$, which determines the range of product lengths a country produces. A large $r$ indicates that a country makes both long and short products, essentially allowing for a large heterogeneity of product lengths. A small $r$ means that there is little room for variation in product lengths, and all products produced will be of approximately the same length. It follows that countries produce products in the range of length $n-r$ to $n$, as $n$ is the maximum product length. The product variety given $r$ is thus given by
\begin{align*} 
d(n,r) = \sum_{s=n-r}^n {n \choose s}, 
\end{align*} 
where ${n \choose s}$ is the number of products of length $s$ that can be made out of $n$ capabilities. 

\begin{figure}[h]
\centering
\includegraphics[width=\linewidth]{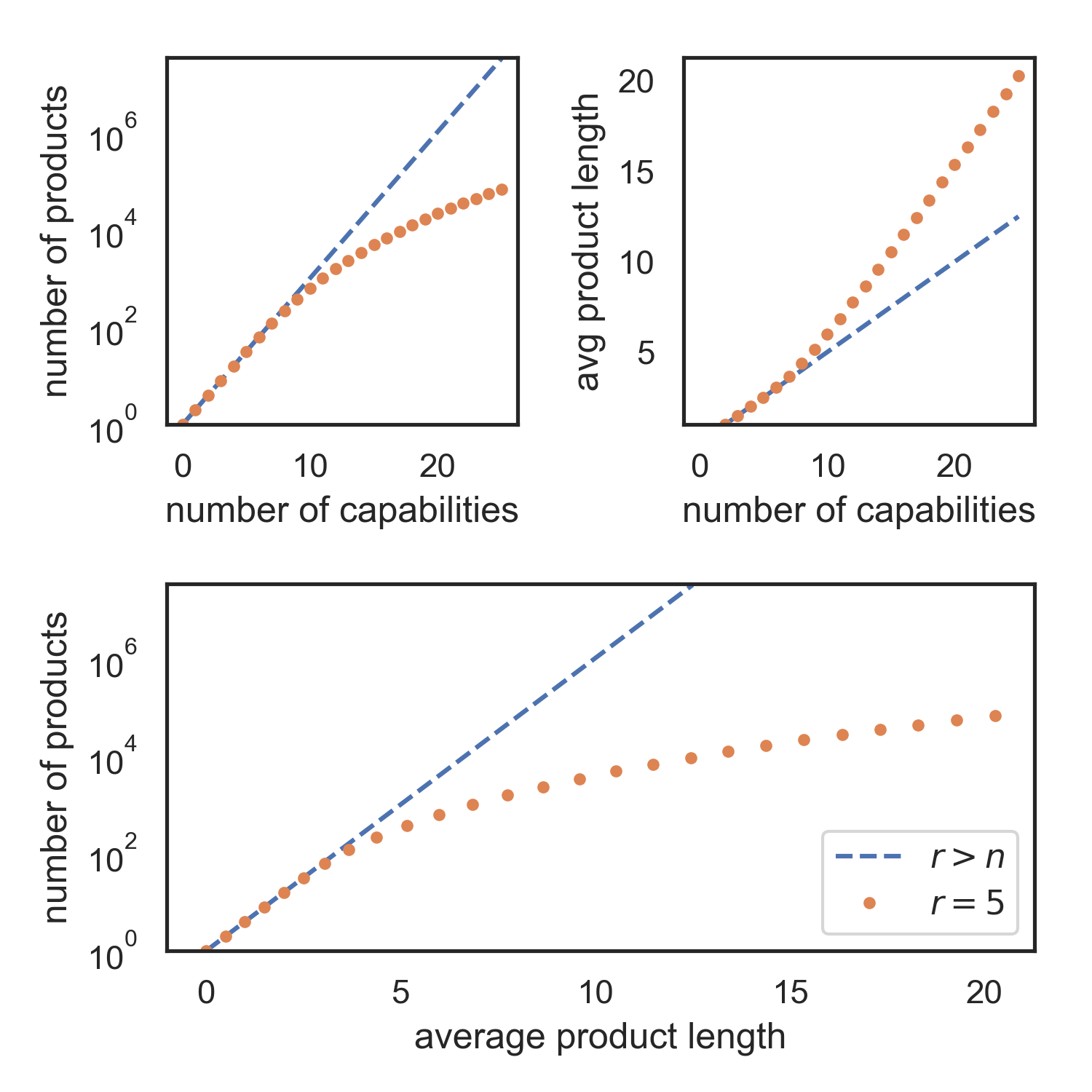}
\caption{The basic dynamics of the model. The blue dashed line shows the case where no products are lost ($r>n$). The orange dots represent the case where $r=5$, so a country only makes products with a length in the range of $n-5$ and $n$. The top left panel shows the relation between product variety $d(n,r)$ and the number of capabilities. For the unconstrained case, there is an exponential relation between the two, showing a linear relationship on a logarithmic scale. Imposing a constraint of $r=5$, the increase in variety slows down as $n$ increases beyond $5$. The top right panel shows average product length $\bar{s}(n)$ as a function of the number of capabilities. In the unconstrained case there is a linear relation, whereas the constrained case shows an acceleration in the increase of average product length once $n>5$ and short products are lost. The bottom panel shows the resulting relation between the product variety and average product length.}
\label{fig:joint1}
\end{figure}

The average product length given $r$ is given by (see \ref{app:avgsr})
\begin{align*} 
\bar{s}(n,r) &= n \frac{d(n-1,r)}{d(n,r)}. 
\end{align*} 
As long as $r \geq n$, the product range forms no constraint on the product lengths and no products are lost. In particular, since $d(n,r) = 2^n$ for $r>n$, we retrieve $\frac{d(n-1,r)}{d(n,r)}= \frac{1}{2}$ so that $\bar{s}(n) = \frac{n}{2}$ as before. When $r<n$, we find that (see \ref{app:limrho1})
\begin{align*} 
\frac{1}{2} < \frac{d(n-1,r)}{d(n,r)} < 1.
\end{align*} 
Assuming that an economy acquires new capabilities one-by-one, the dynamics of the model can then be represented as in Figure \ref{fig:joint1}. Once products starts exiting a country's portfolio, the rate at which the average product length increases in $n$ goes up as the number of capabilities increases, but never exceeds $1$. At the same time, the pace of diversification levels off as more products exit, but a country never loses more products than it gains.

\section{Full model}
\label{S:4}
The assumption that any combination of capabilities leads to a viable product is arguably too strong. More realistically, one may assume that only a fraction of combinations of capabilities result into meaningful products. This can be thought of as imposing a 'recipe book', which describes the combinations of capabilities that are complementary in that they lead to viable products \citep{Hausmann2011, Inoua2016, Fink2017}.

The basic model can be generalized by assuming that every capability is part of a viable product with a given probability $\rho$ \citep{Inoua2016}. Parameter $\rho$ can thus be thought of as reflecting the difficulty to innovation in the sense that not all combinations of capabilities, or 'recipes', lead to viable products. The lower the value of $\rho$, the harder it is to find useful recipes. A combination of $s$ capabilities has probability $\rho^s$ of representing a viable product of length $s$. Hence, it becomes increasingly unlikely that a combination of capabilities leads to a viable product as more capabilities are added, since $\rho^s$ is decreasing in $s$ when $\rho<1$. For $\rho=1$, we recover the basic model described before.

Since there are ${n \choose s}$ possible combinations of $s$ components one can make from the total of $n$ components, and each combination of length $s$ has probability $\rho^s$ of being viable, the expected number of products of length $s$ a country with $n$ components can make is given by $d(n,s) = {n \choose s} \rho^s$. Summing this quantity over all product lengths $s$ gives the expected product variety for a given number of components $n$
\begin{align*}
d(n) = \sum_{s=0}^n {n \choose s} \rho^s = (1+\rho)^n.
\end{align*} 
Since the share of products of length $s$ in a country is given by $\frac{{n \choose s} \rho^s }{d(n)}$, the expected product length given $n$ components can be computed as \citep{Inoua2016}
\begin{align} \label{eq:avgs}
\bar{s}(n) =  \sum_{s=0}^n s \frac{{n \choose s} \rho^s}{d(n)} &=\frac{\rho}{1+\rho}n.
\end{align} 
Note that, as in the basic model before, the expected product length increases linearly with $n$, where the exact rate at which the average product length increases is solely determined by the difficulty parameter $\rho$.

Incorporating the product range using parameter $r$ gives
\begin{align*} 
d(n,r) = \sum_{s=n-r}^n s {n \choose s} \rho^s,
\end{align*} 
and the average product length given $r$ is given by (see \ref{app:avgsr})
\begin{align*} 
\bar{s}(n,r) = \rho n \frac{d(n-1,r)}{d(n,r)}. 
\end{align*} 
In \ref{app:avgsbound} it is shown that for $r<n$, the average product length is bounded from below by  
\begin{align*} 
\frac{\rho}{1+\rho} n < \bar{s}(n,r).
\end{align*} 
Thus once a country starts losing products, the increase of average product length with the number of capabilities starts accelerating. Furthermore, as long as variety is increasing, the increase in average product length is bounded as $\bar{s}(n,r) < \rho n$. 

The model with $\rho<1$ shows an important qualitative difference with the basic model with $\rho=1$, in that for $\rho<1$ a decrease in variety can occur, which happens when more products exit than product enter. The condition for a decline in product variety, i.e. for 'the hump' to occur, is given by (see \ref{app:divr})
\begin{align*}
d(n,r) < {n \choose r} \rho^{n-r-1}.
\end{align*}\\
Once this condition is met, i.e. when a country starts losing more products than it gains, the average product length grows with a rate larger than $\rho$
\begin{align*} 
\bar{s}(n,r) =  \rho n  \frac{d(n-1,r)}{d(n,r)} > \rho n  
\end{align*} 
The model thus predicts that a decrease in variety is accompanied by further acceleration of the increase of the average product length with the number of capabilities as the shortest products are dropped. In this respect, losing the least sophisticated products constitutes a means to improve an economy's complexity.

Finally note that the increasing rate of the average product length is bounded from above by $n$ (see \ref{app:avgsbound}):
\begin{align*} 
\bar{s}(n,r) < n,
\end{align*}
which describes the limiting case in which only the single longest product of length $n$ is produced. 

\begin{figure}[h]
\centering
\includegraphics[width=\linewidth]{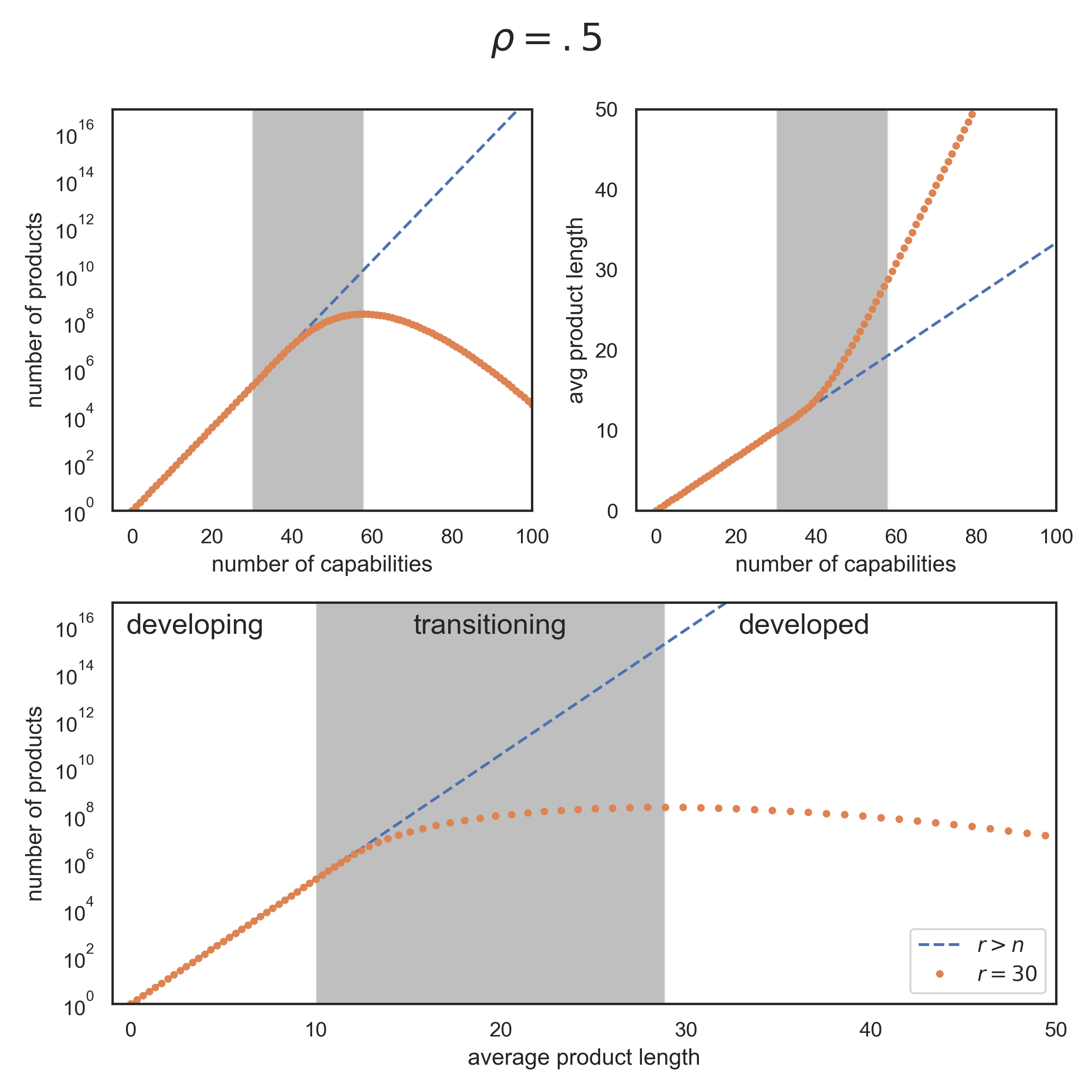}
\caption{The dynamics for the model including a 'recipe book', where every capability is used in a product with probability $\rho=0.5$. The dashed blue line represent the case where there is no constraint by the product range ($r>n$), and the orange dots show the case where $r = 30$. The grey panel's left border indicates the point where $r=n$, and products start to be lost due to the restricted product range. The right edge of the grey panel indicates the location of the 'hump', i.e. when more products exit than products enter. The top left panel shows how the expected product variety increases and then decreases in $n$ for the constrained case. The top right panel shows the expected average product length, which shows an increase in the rate during the transitioning period. The bottom panel shows the relation between product variety and average product length for the three stages of development.} 
\label{fig:joint5}
\end{figure}

In summary, the model exhibits the two 'stages of diversification' as identified empirically by \cite{Imbs2003}, along with a transitory phase in between, known as 'the hump'. An overview of the three stages and their conditions is given in Table \ref{tab:table}. Figure \ref{fig:joint5} further shows the dynamics of the model for the example of $\rho=0.5$. In the model, the first stage of diversification is characterized by an exponential increase in product variety. During this stage no products are lost since the allowed range of product complexities exceeds the total number of capabilities. The average product length increases linearly in $n$ with a rate that is determined by parameter $\rho$. In the transition stage, the simplest products are not produced anymore but the economy is still diversifying, although the rate of diversification slows down. The increases in average product length on the other hand accelerates as the shortest products exit a country's portfolio. In the final stage of diversification, then, more products are lost than gained, so variety decreases as more capabilities are acquired. During this stage, the rate of the average product length further increases and approaches the limiting rate of $1$.

\begin{table}[h]
\centering
\resizebox{.45\textwidth}{!}{
\begin{tabular} { | c | c | c | c | }
    \hline 
     stage & condition & variety & avg. product length \\ 
     \hline  \hline
     developing & $r>n$ &  exponentially increasing & $\bar{s}(n) = \frac{\rho}{1+\rho}n$   \\ 
     \hline 
     transitioning & $r<n$, $d(n,r) > {n \choose r} \rho^{n-r-1}$ & increasing with a decreasing rate & $\frac{\rho}{1+\rho}n \leq \bar{s}(n) \leq \rho n$.  \\ 
     \hline 
     developed & $r<n$, $d(n,r) < {n \choose r} \rho^{n-r-1}$ & decreasing & $\rho n \leq \bar{s}(n) \leq n$.  \\ 
     \hline
\end{tabular}}
\caption{The conditions for the three stages in the model, with the corresponding values of product variety and average product length.  }
\label{tab:table}
\end{table}

\begin{figure}[h]
\centering
\includegraphics[width=\linewidth]{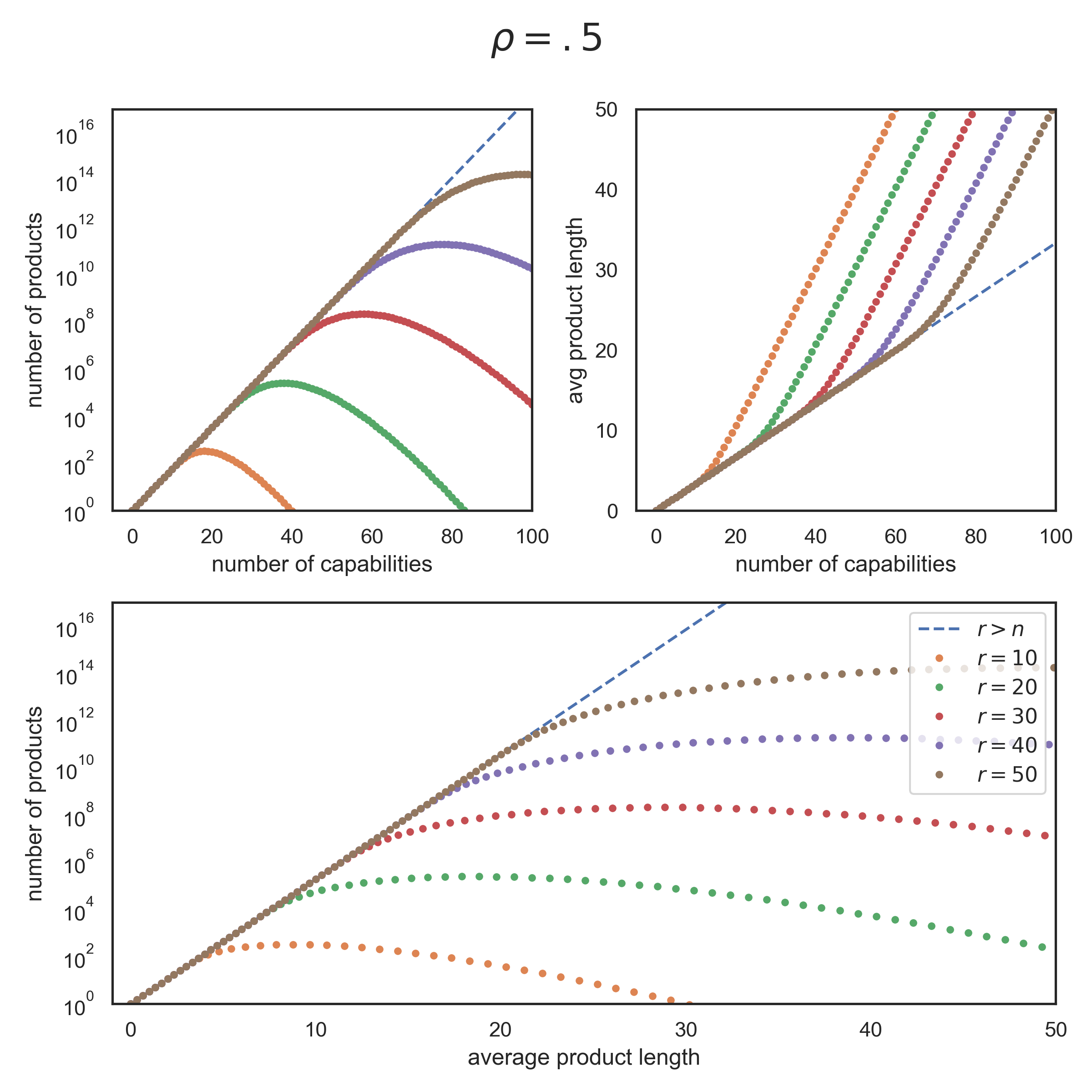}
\caption{The dynamics of the model for different values of product range $r$. The wider the product range (the larger $r$), the later the hump occurs.}
\label{fig:jointr}
\end{figure}

A final feature of the model hold that the product range $r$ determines at what number of capabilities a country enters a new stage of diversification. A country with a large product range $r$ will start losing products at a higher number of capabilities than a country with a small product range. Thus, a large $r$ causes a country to go through the hump later than a country with lower $r$. And, countries with relatively low product range will experience the hump already at a low number of capabilities. The effect of $r$ on the onset of the hump is shown in Figure \ref{fig:jointr}.

\section{Conclusions}
\label{S:5}

%summary + implications 
Elaborating on the combinatorial framework of economic development proposed by \cite{Hausmann2011} and \cite{Inoua2016}, we have modelled an economy as developing over time by acquiring new capabilities one-by-one. Every new capability is recombined with existing capabilities to allow for the production of an ever greater variety of products as well as more complex products. As long as a country produces every product it can produce given its capabilities, variety increases exponentially with the number of capabilities present, while the average product complexity as measured by the required number of capabilities, only increases linearly with the number of capabilities. By assuming that there is a maximum range of product complexities that can be made in a country, one is able to to recover 'the hump' in variety, which refers to the stylized fact that economies first increase and then decrease their product variety as they develop \citep{Imbs2003, Cadot2011}.

%inverstigate r 
The larger the range of product complexities that an economy tolerates to be produced, the longer it takes for the hump to occur. As anticipated by \cite{Imbs2003}, the empirical question that follows holds what country characteristics affect this range, so as to be able to explain why some countries experience the hump earlier in their development than others. For example, the size of a country may be of importance as larger countries may keep on producing low-complexity products for much longer in their low-wage regions compared to small countries where such low-wage regions may be absent. Furthermore, institutional factors including the absence of a minimum wage (prolonging the production of simple products) and trade barriers (preventing the import of simple products from low-wage countries) may further explain a delayed occurrence of the hump.

%Limitations
One objection to the model presented here may regard the way the recipe book is modelled. It was assumed that every capability has an equal chance to be part of any product. It is reasonable to think of some capabilities to be more useful than others in that some capabilities have a higher probability to occur in products than others. Recent empirical research mapping inputs onto outputs has indeed shown that inputs may differ in their prevalence \citep{Fink2017}. The structure of our model does allow for alternative formulations of the recipe book. Note here that for the hump to occur, it only matters whether a recipe book shows a single-peaked distribution of product complexities as the lower bound of the  range of tolerated product complexities will then inevitably pass this peak, as the number of capabilities increases.

\section{Implications}
\label{S:6}
%variety is not complexity 
Turning to the recent burgeoning literature on economic complexity, our model bears an important implication. We have argued that the relationship between product variety (the number of products) and economic complexity (the average number of capabilities used in products) is highly dependent on the stage of development. During an economy's first stage of development, product variety and economic complexity evolve in tandem with variety increasing exponentially and complexity increasing linearly with the number of capabilities. Hence, one could derive an economy's unobservable economic complexity from the logarithm of the observable product variety \citep{Inoua2016}. This relationship, however, changes in a transition stage during which the increase in economic complexity accelerates while the increase in product variety slows down, to eventually reach the final stage of diversification during which product variety even starts declining after going through the hump. As the relationship between variety and complexity depends on the stage of an economy's development, empirical attempts to derive economic complexity from product variety are not grounded theoretically by this model. More precisely, following our model, such attempts may only be meaningful for developing countries being in the first stage of diversification. Note that our theoretical argument to fundamentally distinguish between variety and complexity adds to a recent methodological contribution by \cite{Mealy2018} who disentangle the alleged association between variety and complexity in empirical studies measuring economic complexity.

%support which products are dropped 
One way ahead in empirical research, then, is to collect direct measurements of complexity from observable characteristics of products. For example, and in line with the notion that more complex products are those that require more capabilities to be produced, one could measure a product's complexity from the number of professions involved in its production. In the context of the model we just presented, a direct measure of complexity is important for two reasons. First, using such a measure, one would be able to verify the assertion that economies drop the simplest products from their portfolio, next to other exit determinants as already investigated \citep{Neffke2011, Essletzbichler2015}. While there is some indirect evidence that countries do so \citep{Cadot2011}, we should attempt to verify this empirically. Second, a direct measure of complexity would also be required to further scrutinize the phenomenon of the hump. While our model can replicate the hump as a stylized fact of economic development over time, one is in need of a complexity measure to estimate the exact shape of the hump as a relation between complexity and variety, as predicted by the model.

%GPDpc vs complexity 
The question that remains is how we should understand the relationship between economic development (as understood in terms of stages of diversification) and economic growth (as understood in terms of GDP per capita). We have been able to show that as countries lose the simplest products from their portfolio, they will experience faster increases in the average complexity of the products that remain. One may be tempted, then, to associate the economic complexity of an economy with GDP per capita if one assumes that the average complexity of products in the economy is reflected in the average wage paid to labour. Following this reasoning, our model would predict that the GDP per capita of high-income countries has accelerated over the past decades, while the opposite has been observed. This reasoning, however, suffers from confusing the variety of products in an economy (extensive margin) with their relative shares (intensive margin). To consider the GDP per capita as a proxy for average product complexity would assume that all products have an equal share in the economy as well as in the workforce.

More fundamentally, while economic development can be understood as stemming from the acquisition of new capabilities, there is no reason to believe that new capabilities arrive at a constant rate. For developing countries, the challenge to acquire capabilities may be largely sought in the adoption of capabilities that already exist in the world through channels like imitation, immigration, cooperation and learning. By contrast, countries at the frontier of technological development have to rely on the invention of new capabilities and finding the new combinations with the capabilities they already have \citep{Klinger2004}. Theoretically, then, it is conceivable that the slowdown of growth in high-income countries over the past decades is solely the result of a slowdown in the rate at which new capabilities are acquired. This links to observed changes in the return to R\&D, which arguably underlie to an important extent the acquisition of new capabilities. Indeed, evidence is mounting that the return on R\&D has been  declining over the past decades, precisely because of the difficulty to recombine an ever larger number of knowledge domains into new inventions \citep{Jones2005, Gordon2016}. Yet, to fully appreciate this recent finding in the light of the framework of economic complexity, we are in need of direct measures of capabilities to understand the evolution of economic complexity across space and time.

\section*{Acknowledgments}
Both authors are funded by the Netherlands Organisation for Scientific Research (NWO) under the Vici scheme, number 453-14-014.

%% References

%\clearpage
\section*{References}
\bibliographystyle{model2-names}
\bibliography{library.bib}

\begin{thebibliography}{25}
\expandafter\ifx\csname natexlab\endcsname\relax\def\natexlab#1{#1}\fi
\providecommand{\url}[1]{\texttt{#1}}
\providecommand{\href}[2]{#2}
\providecommand{\path}[1]{#1}
\providecommand{\DOIprefix}{doi:}
\providecommand{\ArXivprefix}{arXiv:}
\providecommand{\URLprefix}{URL: }
\providecommand{\Pubmedprefix}{pmid:}
\providecommand{\doi}[1]{\href{http://dx.doi.org/#1}{\path{#1}}}
\providecommand{\Pubmed}[1]{\href{pmid:#1}{\path{#1}}}
\providecommand{\bibinfo}[2]{#2}
\ifx\xfnm\relax \def\xfnm[#1]{\unskip,\space#1}\fi
%Type = Article
\bibitem[{Al-Marhubi(2000)}]{Al-Marhubi2000}
\bibinfo{author}{Al-Marhubi, F.}, \bibinfo{year}{2000}.
\newblock \bibinfo{title}{{Export diversification and growth: an empirical
  investigation}}.
\newblock \bibinfo{journal}{Applied Economics Letters}
\newblock \bibinfo{volume}{7}, \bibinfo{pages}{559--562},
  \DOIprefix\doi{10.1080/13504850050059005}.
%Type = Article
\bibitem[{Brummitt et~al.(2017)Brummitt, Huremovi{\'{c}}, Pin, Bonds and
  Vega-Redondo}]{Brummitt2017}
\bibinfo{author}{Brummitt, C.D.}, \bibinfo{author}{Huremovi{\'{c}}, K.},
  \bibinfo{author}{Pin, P.}, \bibinfo{author}{Bonds, M.H.},
  \bibinfo{author}{Vega-Redondo, F.}, \bibinfo{year}{2017}.
\newblock \bibinfo{title}{{Contagious disruptions and complexity traps in
  economic development}}.
\newblock \bibinfo{journal}{Nature Human Behaviour}
\newblock \bibinfo{volume}{1}, \bibinfo{pages}{665--672},
  \DOIprefix\doi{10.1038/s41562-017-0190-6}.
%Type = Article
\bibitem[{Cadot et~al.(2011)Cadot, Carr{\`{e}}re and Strauss-Kahn}]{Cadot2011}
\bibinfo{author}{Cadot, O.}, \bibinfo{author}{Carr{\`{e}}re, C.},
  \bibinfo{author}{Strauss-Kahn, V.}, \bibinfo{year}{2011}.
\newblock \bibinfo{title}{{Export Diversification: What's behind the Hump?}}
\newblock \bibinfo{journal}{Review of Economics and Statistics}
\newblock \bibinfo{volume}{93}, \bibinfo{pages}{590--605},
  \DOIprefix\doi{10.1162/REST{\_}a{\_}00078}.
%Type = Article
\bibitem[{Cristelli et~al.(2015)Cristelli, Tacchella and
  Pietronero}]{Cristelli2015}
\bibinfo{author}{Cristelli, M.}, \bibinfo{author}{Tacchella, A.},
  \bibinfo{author}{Pietronero, L.}, \bibinfo{year}{2015}.
\newblock \bibinfo{title}{{The Heterogeneous Dynamics of Economic Complexity}}.
\newblock \bibinfo{journal}{PLOS ONE}
\newblock \bibinfo{volume}{10}, \bibinfo{pages}{e0117174},
  \DOIprefix\doi{10.1371/journal.pone.0117174}.
%Type = Article
\bibitem[{Essletzbichler(2015)}]{Essletzbichler2015}
\bibinfo{author}{Essletzbichler, J.}, \bibinfo{year}{2015}.
\newblock \bibinfo{title}{{Relatedness, Industrial Branching and Technological
  Cohesion in US Metropolitan Areas}}.
\newblock \bibinfo{journal}{Regional Studies}
\newblock \bibinfo{volume}{49}, \bibinfo{pages}{752--766},
  \DOIprefix\doi{10.1080/00343404.2013.806793}.
%Type = Article
\bibitem[{Fink et~al.(2017)Fink, Reeves, Palma and Farr}]{Fink2017}
\bibinfo{author}{Fink, T.M.A.}, \bibinfo{author}{Reeves, M.},
  \bibinfo{author}{Palma, R.}, \bibinfo{author}{Farr, R.S.},
  \bibinfo{year}{2017}.
\newblock \bibinfo{title}{{Serendipity and strategy in rapid innovation}}.
\newblock \bibinfo{journal}{Nature Communications}
\newblock \bibinfo{volume}{8}, \bibinfo{pages}{2002},
  \DOIprefix\doi{10.1038/s41467-017-02042-w}.
%Type = Book
\bibitem[{Gordon(2016)}]{Gordon2016}
\bibinfo{author}{Gordon, R.J.}, \bibinfo{year}{2016}.
\newblock \bibinfo{title}{{The Rise and Fall of American Growth}}.
\newblock
\newblock \bibinfo{publisher}{Princeton University Press},
  \bibinfo{address}{Princeton}, \DOIprefix\doi{10.1515/9781400873302}.
%Type = Article
\bibitem[{Hausmann and Hidalgo(2011)}]{Hausmann2011}
\bibinfo{author}{Hausmann, R.}, \bibinfo{author}{Hidalgo, C.A.},
  \bibinfo{year}{2011}.
\newblock \bibinfo{title}{{The network structure of economic output}}.
\newblock \bibinfo{journal}{Journal of Economic Growth}
\newblock \bibinfo{volume}{16}, \bibinfo{pages}{309--342},
  \DOIprefix\doi{10.1007/s10887-011-9071-4}.
%Type = Article
\bibitem[{Hausmann et~al.(2007)Hausmann, Hwang and Rodrik}]{Hausmann2007}
\bibinfo{author}{Hausmann, R.}, \bibinfo{author}{Hwang, J.},
  \bibinfo{author}{Rodrik, D.}, \bibinfo{year}{2007}.
\newblock \bibinfo{title}{{What you export matters}}.
\newblock \bibinfo{journal}{Journal of Economic Growth}
\newblock \bibinfo{volume}{12}, \bibinfo{pages}{1--25},
  \DOIprefix\doi{10.1007/s10887-006-9009-4}.
%Type = Article
\bibitem[{Herzer and Nowak-Lehnmann(2006)}]{Herzer2006}
\bibinfo{author}{Herzer, D.}, \bibinfo{author}{Nowak-Lehnmann, F.D.},
  \bibinfo{year}{2006}.
\newblock \bibinfo{title}{{What does export diversification do for growth? An
  econometric analysis}}.
\newblock \bibinfo{journal}{Applied Economics}
\newblock \bibinfo{volume}{38}, \bibinfo{pages}{1825--1838},
  \DOIprefix\doi{10.1080/00036840500426983}.
%Type = Article
\bibitem[{Hesse(2008)}]{Hesse2008}
\bibinfo{author}{Hesse, H.}, \bibinfo{year}{2008}.
\newblock \bibinfo{title}{{Export Diversification and Economic Growth}}.
\newblock \bibinfo{journal}{Commission on Growth and Development Working Paper}
\newblock \bibinfo{volume}{21}.
%Type = Article
\bibitem[{Hidalgo and Hausmann(2009)}]{Hidalgo2009}
\bibinfo{author}{Hidalgo, C.A.}, \bibinfo{author}{Hausmann, R.},
  \bibinfo{year}{2009}.
\newblock \bibinfo{title}{{The building blocks of economic complexity}}.
\newblock \bibinfo{journal}{Proceedings of the National Academy of Sciences}
\newblock \bibinfo{volume}{106}, \bibinfo{pages}{10570--10575},
  \DOIprefix\doi{10.1073/pnas.0900943106}.
%Type = Article
\bibitem[{Hidalgo et~al.(2007)Hidalgo, Klinger, Barabasi and
  Hausmann}]{Hidalgo2007}
\bibinfo{author}{Hidalgo, C.A.}, \bibinfo{author}{Klinger, B.},
  \bibinfo{author}{Barabasi, A.L.}, \bibinfo{author}{Hausmann, R.},
  \bibinfo{year}{2007}.
\newblock \bibinfo{title}{{The Product Space Conditions the Development of
  Nations}}.
\newblock \bibinfo{journal}{Science}
\newblock \bibinfo{volume}{317}, \bibinfo{pages}{482--487},
  \DOIprefix\doi{10.1126/science.1144581}.
%Type = Article
\bibitem[{Imbs and Wacziarg(2003)}]{Imbs2003}
\bibinfo{author}{Imbs, J.}, \bibinfo{author}{Wacziarg, R.},
  \bibinfo{year}{2003}.
\newblock \bibinfo{title}{{Stages of diversification}}.
\newblock \bibinfo{journal}{American Economic Review}
\newblock \bibinfo{volume}{93}, \bibinfo{pages}{63--86},
  \DOIprefix\doi{10.1257/000282803321455160}.
%Type = Article
\bibitem[{Inoua(2016)}]{Inoua2016}
\bibinfo{author}{Inoua, S.}, \bibinfo{year}{2016}.
\newblock \bibinfo{title}{{A Simple Measure of Economic Complexity}}.
\newblock \bibinfo{journal}{http://arxiv.org/abs/1601.05012} .
%Type = Article
\bibitem[{Jones(2009)}]{Jones2005}
\bibinfo{author}{Jones, B.F.}, \bibinfo{year}{2009}.
\newblock \bibinfo{title}{{The Burden of Knowledge and the “Death of the
  Renaissance Man”: Is Innovation Getting Harder?}}
\newblock \bibinfo{journal}{Review of Economic Studies}
\newblock \bibinfo{volume}{76}, \bibinfo{pages}{283--317},
  \DOIprefix\doi{10.1111/j.1467-937X.2008.00531.x}.
%Type = Book
\bibitem[{Kauffman(1993)}]{Kauffman1993}
\bibinfo{author}{Kauffman, S.A.}, \bibinfo{year}{1993}.
\newblock \bibinfo{title}{{The origins of order: self-organization and
  selection in evolution}}.
\newblock
\newblock \bibinfo{publisher}{Oxford University Press}.
%Type = Article
\bibitem[{Klinger and Lederman(2004)}]{Klinger2004}
\bibinfo{author}{Klinger, B.}, \bibinfo{author}{Lederman, D.},
  \bibinfo{year}{2004}.
\newblock \bibinfo{title}{{Diversification, Innovation, and Imitation inside
  the Global Technological Frontier}}.
\newblock \bibinfo{journal}{World Bank Policy Research Working Paper No .3872}
  .
%Type = Article
\bibitem[{Kremer(1993)}]{Kremer1993}
\bibinfo{author}{Kremer, M.}, \bibinfo{year}{1993}.
\newblock \bibinfo{title}{{The O-Ring Theory of Economic Development}}.
\newblock \bibinfo{journal}{The Quarterly Journal of Economics}
\newblock \bibinfo{volume}{108}, \bibinfo{pages}{551--575},
  \DOIprefix\doi{10.2307/2118400}.
%Type = Article
\bibitem[{Lall(2000)}]{Lall2000}
\bibinfo{author}{Lall, S.}, \bibinfo{year}{2000}.
\newblock \bibinfo{title}{{The technological structure and performance of
  developing country manufactured exports, 1985-98}}.
\newblock \bibinfo{journal}{Oxford Development Studies}
\newblock \bibinfo{volume}{28}, \bibinfo{pages}{337--369},
  \DOIprefix\doi{10.1080/713688318}.
%Type = Article
\bibitem[{Lall et~al.(2006)Lall, Weiss and Zhang}]{Lall2006}
\bibinfo{author}{Lall, S.}, \bibinfo{author}{Weiss, J.},
  \bibinfo{author}{Zhang, J.}, \bibinfo{year}{2006}.
\newblock \bibinfo{title}{{The "sophistication" of exports: A new trade
  measure}}.
\newblock \bibinfo{journal}{World Development}
\newblock \bibinfo{volume}{34}, \bibinfo{pages}{222--237},
  \DOIprefix\doi{10.1016/j.worlddev.2005.09.002}.
%Type = Article
\bibitem[{Mealy et~al.(2018)Mealy, Farmer and Teytelboym}]{Mealy2018}
\bibinfo{author}{Mealy, P.}, \bibinfo{author}{Farmer, J.D.},
  \bibinfo{author}{Teytelboym, A.}, \bibinfo{year}{2018}.
\newblock \bibinfo{title}{{A New Interpretation of the Economic Complexity
  Index}} , \bibinfo{pages}{1--36}, \DOIprefix\doi{10.2139/ssrn.3075591}.
%Type = Article
\bibitem[{Neffke et~al.(2011)Neffke, Henning and Boschma}]{Neffke2011}
\bibinfo{author}{Neffke, F.}, \bibinfo{author}{Henning, M.},
  \bibinfo{author}{Boschma, R.}, \bibinfo{year}{2011}.
\newblock \bibinfo{title}{{How Do Regions Diversify over Time? Industry
  Relatedness and the Development of New Growth Paths in Regions}}.
\newblock \bibinfo{journal}{Economic Geography}
\newblock \bibinfo{volume}{87}, \bibinfo{pages}{237--265},
  \DOIprefix\doi{10.1111/j.1944-8287.2011.01121.x}.
%Type = Article
\bibitem[{Sutton and Trefler(2016)}]{Sutton2016}
\bibinfo{author}{Sutton, J.}, \bibinfo{author}{Trefler, D.},
  \bibinfo{year}{2016}.
\newblock \bibinfo{title}{{Capabilities , Wealth , and Trade}}.
\newblock \bibinfo{journal}{Journal of Political Economy}
\newblock \bibinfo{volume}{124}, \bibinfo{pages}{826--878}.
%Type = Article
\bibitem[{Tacchella et~al.(2012)Tacchella, Cristelli, Caldarelli, Gabrielli and
  Pietronero}]{Tacchella2012}
\bibinfo{author}{Tacchella, A.}, \bibinfo{author}{Cristelli, M.},
  \bibinfo{author}{Caldarelli, G.}, \bibinfo{author}{Gabrielli, A.},
  \bibinfo{author}{Pietronero, L.}, \bibinfo{year}{2012}.
\newblock \bibinfo{title}{{A New Metrics for Countries' Fitness and Products'
  Complexity}}.
\newblock \bibinfo{journal}{Scientific Reports}
\newblock \bibinfo{volume}{2}, \bibinfo{pages}{723},
  \DOIprefix\doi{10.1038/srep00723}.

\end{thebibliography}






% \end{thebibliography}

\appendix

\section{Derivations of full model quantities} 

\subsection{Average product length}
\label{app:avgs}
\begin{align*} 
\bar{s}(n) =  \sum_{s=0}^n s \frac{d(n,s)}{d(n)} &= \frac{\sum_{s=0}^n s {n \choose s} \rho^s}{(1+\rho)^n}  \nonumber \\
&=(1+\rho)^{-n} \sum_{s=1}^n s \frac{n}{s} {n-1\choose s-1}\rho^s  \nonumber\\
&= \rho (1+\rho)^{-n} n \sum_{s=1}^n {n-1\choose s-1}\rho^{s-1}  \nonumber\\
&= \rho (1+\rho)^{-n} n \sum_{x=0}^{n-1} {n-1\choose x}\rho^{x}  \nonumber\\
&= \rho (1+\rho)^{-n} n (1+\rho)^{n-1}\nonumber \\
&=\frac{\rho}{1+\rho}n.
\end{align*} 

\subsection{Bounds on average product length for $\rho=1$}      \label{app:limrho1} 
For $\rho=1$, we have that 
\begin{align*} 
\frac{1}{2} \leq \frac{d(n-1,r)}{d(n,r)} < 1. 
\end{align*} 

First, note that 
\begin{align}\label{eq:diff}
d(n,r) = 2d(n-1,r) - {n-1 \choose r}
\end{align} 
by \ref{app:divr}. Now since 
\begin{align*} 
d(n-1,r) = \sum_{s=n-1-r}^{n-1} {n-1 \choose s} &=\sum_{s=n-r}^{n-1} {n-1 \choose s} + {n-1 \choose n-1-r}> {n-1 \choose n-1-r},
\end{align*}
we have that 
\begin{align*}
d(n,r) &= 2d(n-1,r) - {n-1 \choose r}\\
&= 2d(n-1,r) - {n-1 \choose n-1-r}\\
&> 2d(n-1,r) - d(n-1,r)\\
&> d(n-1,r),
\end{align*} 
so that $\frac{d(n-1,r)}{d(n,r)}<1$.\\

Furthermore, since ${n-1 \choose r}>0$, \eqref{eq:diff} gives that 
\begin{align*} 
d(n,r) < 2 d(n-1,r)
\end{align*}
and thus $\frac{d(n-1,r)}{d(n,r)}> \frac{1}{2}$ for $r<n$.

\subsection{Average product length given $r$}
\label{app:avgsr}
First, note that the total product length is given by 
\begin{align*} 
d(n,r)\bar{s}(n,r) &= \sum_{s=n-r}^n s {n \choose s} \rho^s \\
&=n\rho \sum_{s=n-r}^{n} \frac{(n-1)!}{(s-1)! (n-s)!} \rho^{s-1} \\
&=n \rho \sum_{s'=n-r-1}^{n-1} \frac{(n-1)!}{s'! (n-s'-1)!}\rho^{s'} \\
&=n \rho \sum_{s'=n-r-1}^{n-1} {n-1 \choose s'}\rho^{s'}\\
&= n\rho  d(n-1,r) 
\end{align*}
so that the average product length is given by
\begin{align*} 
\bar{s}(n,r) = n\rho \frac{d(n-1,r)}{d(n,r)}.
\end{align*}

\subsection{Diversification including $r$}
\label{app:divr}
The grows in product variety is given by 
\begin{align*} 
d(n+1,r) - d(n,r) &= \sum_{s=n+1-r}^{n+1} {n+1 \choose s} \rho^s - \sum_{s=n-r}^n {n \choose s} \rho^s\\ 
&= \sum_{s=n+1-r}^{n+1} {n+1 \choose s} \rho^s - \sum_{s=n-r}^{n+1} \frac{n+1-s}{n+1} {n+1 \choose s} \rho^s\\
&= \sum_{s=n+1-r}^{n+1} {n+1 \choose s} \rho^s - \sum_{s=n+1-r}^{n+1} \frac{n+1-s}{n+1} {n+1 \choose s} \rho^s - \frac{r+1}{n+1} {n+1 \choose n-r} \rho^{n-r}\\
&= \sum_{s=n+1-r}^{n+1} \frac{s}{n+1} {n+1 \choose s} \rho^s -
{n \choose r} \rho^{n-r}\\
&= \rho d(n,r)  - {n \choose r} \rho^{n-r}. 
\end{align*} 
Hence variety starts decreasing for $n,r$ when 
\begin{align*}
d(n,r)  < {n \choose r} \rho^{n-r-1}.
\end{align*} 

\subsection{Bounds on average word length}  \label{app:avgsbound}
We show that 
\begin{align*} 
\frac{\rho}{1+\rho} n <\bar{s}(n,r) < n. 
\end{align*} 

From \ref{app:divr} we have 
\begin{align*} 
d(n,r) = (1+\rho) d(n-1,r) - {n-1 \choose r} \rho^{n-r-1}, 
\end{align*} 
so that 
\begin{align*} 
\frac{d(n-1,r)}{d(n,r)} &= \frac{d(n-1,r)}{(1+\rho)d(n-1,r)- {n-1 \choose r} \rho^{n-r-1}}\\
&> \frac{d(n-1,r)}{(1+\rho)d(n-1,r)}= \frac{1}{1+\rho}.
\end{align*}  
This means that a lower bound on average word length is given by  
\begin{align*} 
\bar{s}(n,r) = n \rho \frac{d(n-1,r)}{d(n,r)} > n \frac{\rho}{1+\rho}
\end{align*} 
since ${n-1 \choose r} \rho^{n-r-1}>0$. \\

To find an upper bound, first note that 
\begin{align*} 
{n-1 \choose r} \rho^{n-r-1} < \sum_{s=n-r-1}^{n-1} {n \choose s} \rho^s = d(n-1,r).
\end{align*} 
We then find that
\begin{align*} 
d(n,r) &= (1+\rho) d(n-1,r) - {n-1 \choose r} \rho^{n-r-1}\\
&= (1+\rho) d(n-1,r) - {n-1 \choose n-1-r} \rho^{n-r-1}\\
&> (1+\rho) d(n-1,r) - d(n-1,r)\\
&> \rho d(n-1,r),
\end{align*} 
so that 
\begin{align*} 
\frac{d(n-1,r)}{d(n,r)} < \frac{1}{\rho},
\end{align*} 
and 
\begin{align*} 
\bar{s}(n,r) = n \rho \frac{d(n-1,r)}{d(n,r)} < n. 
\end{align*}

\end{document}